\documentclass{article}

\usepackage{PRIMEarxiv}

\usepackage[utf8]{inputenc} 
\usepackage[T1]{fontenc}    
\usepackage{hyperref}       
\usepackage{url}            
\usepackage{booktabs}       
\usepackage{amsfonts}       
\usepackage{nicefrac}       
\usepackage{microtype}      
\usepackage{lipsum}
\usepackage{fancyhdr}       
\usepackage{graphicx}       
\graphicspath{{media/}}     

\title{Short- and long-term relationships between the Yucatan Channel transport and the Loop Current System
\thanks{\textit{\underline{Citation}}: 
\textbf{Authors. Title. Pages.... DOI:000000/11111.}} 
}

\author{
  Efra\'in Moreles \\
  Instituto de Ciencias del Mar y Limnolog\'ia, Unidad Acad\'emica Procesos Oce\'anicos y Costeros \\
  Universidad Nacional Aut\'onoma de M\'exico \\
  04510, Coyoac\'an, Mexico City, Mexico\\
  \texttt{moreles@cmarl.unam.mx} \\
   \AND
   Benjam\'in Mart\'inez-L\'opez \\
   Instituto de Ciencias de la Atm\'osfera y Cambio Clim\'atico, Universidad Nacional Aut\'onoma de M\'exico \\
   04510, Coyoac\'an, Mexico City, Mexico \\
   \And
   Susana Higuera-Parra \\
   Departamento de Observaci\'on y Estudio de la Tierra, la Atm\'osfera y el Oc\'eano, El Colegio de la Frontera Sur \\
   77014, Chetumal, Quintana Roo, Mexico \\
   \And
   Erick R. Olvera-Prado \\
   Departamento de Ingenier\'ia de Procesos e Hidr\'aulica, CBI, Universidad Aut\'onoma Metropolitana-Iztapalapa \\
   09340, Iztapalapa, Mexico City, Mexico \\
  \And
   Jorge Zavala-Hidalgo \\
   Instituto de Ciencias de la Atm\'osfera y Cambio Clim\'atico, Universidad Nacional Aut\'onoma de M\'exico \\
   04510, Coyoac\'an, Mexico City, Mexico \\
}

\begin{document}
\maketitle

\begin{abstract}
This work uses twin 22-year free-running simulations of the Gulf of Mexico hydrodynamics performed with the HYCOM, one considering only ocean dynamics and the other incorporating atmospheric forcing, to study the behavior of the Yucatan Channel transport (YCT), the Loop Current (LC), the Loop Current Eddies (LCEs), their relationships, and the atmospheric forcing effect on them in short (daily) and long (monthly) time scales. A more comprehensive description of the LC intrusion and LCE separations was obtained by considering the upper eastern or western YCT (whose magnitudes are determined by the longitudinal displacements of the Yucatan Current's core), a perspective not evident when considering the upper total YCT; specifically, the eastern YCT provides the most meaningful description of the studied processes. Atmospheric forcing mainly affects the extended stage of the LC by creating a higher dispersion in the YCT and LC circulation values in comparison when considering only ocean dynamics. For the long-term analysis, standardized indexes that integrate the daily values of the eastern YCT and LC circulation in time were used; their temporal propagation and persistence (the changes of their characteristics from short to long time scales) were studied. Intrinsic ocean dynamics produces a persistent YCT and LC intrusion behavior and consistent LCE separation patterns from daily to 5-month scales. The atmospheric forcing effects are more emphasized on the LC intrusion and LCE separations than on the YCT: the YCT persistence is maintained but not that of the LC intrusion. An increased occurrence of LCE separations with low or moderate LC intrusion is expected due to climate change. Using the standardized indexes of the LC metrics to construct a predictive model of the LC intrusion and LCE separations using only current and past LC information is proposed for future research.
\end{abstract}


\section{Introduction}
\begin{figure}[h]
	\centering
		\includegraphics[angle=0, width=0.95\textwidth]{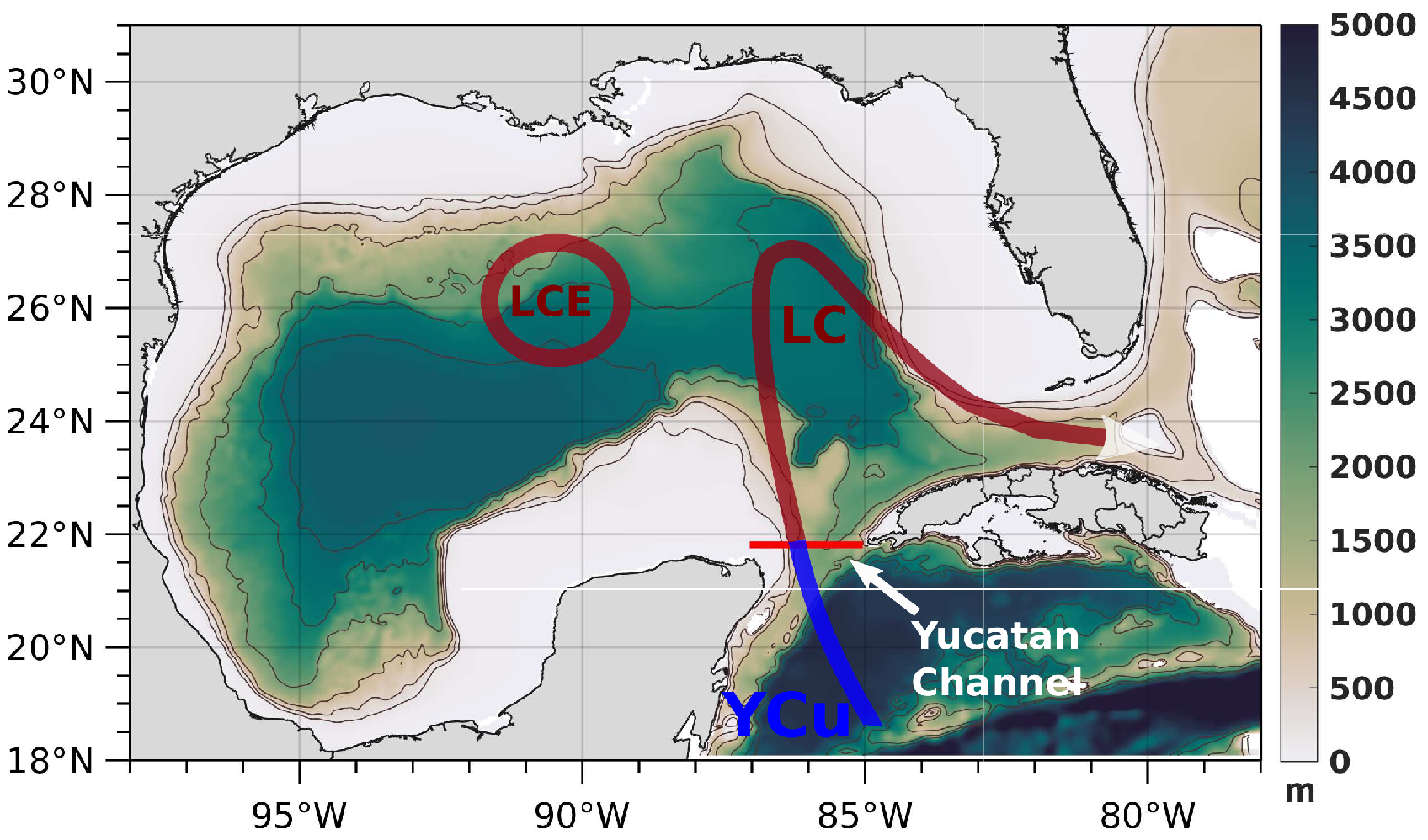}
		\caption{Map of the Gulf of Mexico and a representation of its main circulation features: the Yucatan Current (YCu), the Loop Current (LC), and the Loop Current Eddies (LCEs). The red line at $21.8^{\circ}\mathrm{N}$ indicates the transect used to calcute the water flux through the Yucatan Channel. The black contours represent the 200, 500, 1000, 2000, 3000, and 3500 m isobaths.}
		\label{Fig1}
\end{figure}

The Gulf of Mexico is a sea of prime importance in the North Atlantic Ocean (Fig~\ref{Fig1}), with a significant role in defining the weather and climate of the region \cite{Muller2015}. The main ocean circulation pattern in the Gulf is the Loop Current System (LCS), composed of the Loop Current (LC) and the Loop Current Eddies (LCEs), the large anticyclonic eddies that detach from it \cite{Schmitz2005, Sturges2000, NASEM2018}. The LC originates in the Yucatan Channel as the Yucatan Current (YCu) and penetrates in the Gulf, developing in a continuum of stages between the retracted and extended stages \cite{Leben2005, NASEM2018, Oey2005, Schmitz2005}.

A thorough understanding of the Gulf ocean dynamics requires a comprehensive study of the LCS dynamics, specifically the LC intrusion and LCE separation events. The LCS dynamics is a complex process involving numerous phenomena (e.g., the state of the ocean, atmospheric forcing, and mesoscale eddies in the northern Gulf) interacting highly nonlinearly  \cite{Hurlburt1980, Schmitz2005, NASEM2018, Athie2020, Moreles2021, Higuera2023, Yang2023}. Among all the different phenomena involved in the LCS dynamics, the Yucatan Channel transport (YCT), the water flux through the Yucatan Channel, seems particularly relevant to comprehending and predicting the LC intrusion and LCE separation events \cite{Hurlburt1980, Hurlburt1982, Sheinbaum2002, Oey2003, Ezer2003, Lin2009, LeHenaff2012, Chang2012, Chang2013, Athie2015, Athie2020, Candela2019, Moreles2021, Manta2023, Yang2023}.

Ezer et al. \cite{Ezer2003} found that LCE detachments are usually preceded by the increase in the YCT and occur when it is near its peak. Le H\'enaff  et al. \cite{LeHenaff2012} found that an enhanced YCT generates an increase in the LCE separation period, while its diminishment leads to a more extended and active LC. Moreles et al. \cite{Moreles2021} found significant relationships between the YCT, LC intrusion, and the diameter and separation period of the LCEs. The primary mechanism contributing to the YCT variability is the longitudinal displacements of the YCu, which have been associated with perturbations (eddies) coming from the Caribbean Sea and the subsequent LCE detachment \cite{Abascal2003, Candela2003, Sheinbaum2016, Athie2012, Athie2015, Athie2020} and with the interaction of the LC with mesoscale eddies in the northern Gulf \cite{Higuera2023, Yang2023}.

While previous studies have provided valuable insights into the relationships between the YCT and LCS, their focus has been primarily on the short term (e.g., daily time scales). Given that the YCT evolves on longer time scales than the LCS, their behavior, relationships, and persistence will likely differ at different time scales. The temporal propagation of the YCT and LCS characteristics (the changes of their characteristics from short to long time scales) will depend on the state of the ocean and its forcing. Studies of the relationships between the YCT and the LCS in the long term (monthly time scales) have yet to be carried out; furthermore, there is a great interest in exploring the impact of atmospheric forcing on the YCT and LCS dynamics \cite{Oey2003, Hall2016, Athie2020, Moreles2021, Higuera2023}. Many questions remain concerning the long-term behavior of the YCT and LCS, and long-term and more realistic studies are needed to obtain conclusive results.

Studies addressing the topics mentioned above will improve our comprehension of the Gulf dynamics by providing relevant information regarding the YCT and LCS, such as the persistence and changes in their dynamical relationships at different time scales, that can be directly applied in predictive models. This work investigates the short- and long-term behavior (from daily to monthly time scales) of the YCT, the LC intrusion, the LCE separations, the relationships between them, and how atmospheric forcing modifies them. This study focuses on identifying the relationships between YCT and LCS at different time scales; it does not address the mechanisms causing variations in them.

In order to achieve the objectives, long-term free-running simulations of the Gulf hydrodynamics performed with the HYbrid Coordinate Ocean Model (HYCOM) were used. Twin simulations were used: one considering only ocean dynamics and the other incorporating atmospheric forcing (mass, momentum, and energy fluxes between the ocean and the atmosphere). The daily time series of the upper YCT and the LC metrics were calculated; then, their corresponding time series in the long term were constructed using a standard methodology.

\section{Methods}

\subsection{Data}

This study employed daily outputs of twin simulations of the Gulf hydrodynamics performed with the HYCOM: one focused solely on ocean dynamics, while the other incorporated atmospheric forcing (mass, momentum, and energy fluxes between the ocean and the atmosphere). The simulations are the same as those used by Higuera-Parra et al. \cite{Higuera2023} and Olvera-Prado et al. \cite{Olvera2023}. The simulation considering only ocean dynamics is referred to as the experiment of intrinsic ocean dynamics and denoted as experiment NoAF and the simulation incorporating atmospheric forcing is denoted as experiment AF; the description of these simulations can be found in Higuera-Parra et al. \cite{Higuera2023}.

\subsection{Yucatan Channel transport}

The YCT is commonly associated with the transport of the YCu, a current confined to the west of $85.6^{\circ}\mathrm{W}$ and above approximately 800 m depth \cite{Athie2015}. Studies addressing the relationship between the YCT and the LCS should distinguish between the total, western, or eastern YCT since each contains different dynamic information \cite{Athie2020}. The western and eastern transports have an inverse or almost compensating behavior at different time scales as found by Athi\'e et al. \cite{Athie2020}; these authors found a significant but weak correlation between the LC intrusion and the western transport during 2008-2016 (higher for specific periods) and a nonsignificant correlation when considering the total transport. A relevant result of Athi\'e et al. \cite{Athie2020} is that the eastern transport is not just passively responding to the western transport; the eastern transport has a characteristic dynamics and contributes to the west-east asymmetry in the Yucatan Channel flow. 

Based on the results of Athi\'e et al. \cite{Athie2020}, this study explored the relationships between the LCS and the total, west, and east YCT, considering a Yucatan Channel division longitude at $86.0^{\circ}\mathrm{W}$. The YCT was computed along a section at $21.8^{\circ}\mathrm{N}$ (the red line in Fig~\ref{Fig1}), which was selected to represent the YCT variability; since approximately 90\% of the total YCT occurs in the first 500 m depth \cite{Candela2019, Rousset2010, Rousset2011, Sheinbaum2002}, the vertically integrated meridional YCT above 500 m depth was employed. This way, daily time series of the upper YCT in the total section ($\mathrm{YCT_{T}}$), the west section ($\mathrm{YCT_{W}}$), and the east section ($\mathrm{YCT_{E}}$) were obtained.

\subsection{Loop Current intrusion}

For the description and time monitoring of the LCS, the LC metrics proposed by Hamilton et al. \cite{Hamilton2000} were employed, which identify the LC location in time. From the time series of the LC metrics, the LC intrusion is identified by a sustained increase in them, whereas the LCE detachment and separation events are identified with the local minima in them after an extended LC intrusion. Then, the shedding period and diameter of the LCEs can be calculated. According to Hamilton et al. \cite{Hamilton2000}, the LC circulation ($\mathrm{LC_{C}}$) is a dynamical metric and the only one among all LC metrics with the most consistent behavior in identifying LCE detachments and separations. In this work, the daily $\mathrm{LC_{C}}$ was used, and only the LCE separations (LCE detachments not followed by a reattachment to the mean LC flow) were considered.

\subsection{Long term behavior}

This study takes a new approach to analyzing the long-term behavior of the YCT and $\mathrm{LC_{C}}$. Standardized indexes that integrate the time behavior of their associated variables at different scales were constructed following the methodology proposed by McKee et al. \cite{McKee1993}, who created the standardized precipitation index to analyze droughts, including various of their characteristics: their onset and their propagation and persistence at different time scales. Since its creation, this methodology has been widely used in numerous subsequent studies of droughts and has been the basis for developing more specialized methodologies.

By applying the methodology of McKee et al. \cite{McKee1993}, standardized indexes for the $\mathrm{YCT}$ ($\mathrm{SI_{YCT}}$) and $\mathrm{LC_{C}}$ ($\mathrm{SI_{LC}}$) were obtained. This study aims to shed light on their long-term behavior; thus, standardized indexes for time scales of one, three, and five months were calculated. Similar to the studies employing the standardized precipitation index to analyze droughts (cf. \cite{McKee1993}), this study considers that the standardized indexes $\mathrm{SI_{YCT}}$ and $\mathrm{SI_{LC}}$ are adequate for studying the long-term behavior of the YCT and LC intrusion and describing their propagation and persistence at different time scales.

\section{Results}

\subsection{Short-term behavior of the YCT and LCS}

\begin{figure}[p]
	\centering
		\includegraphics[angle=0, width=0.95\textwidth]{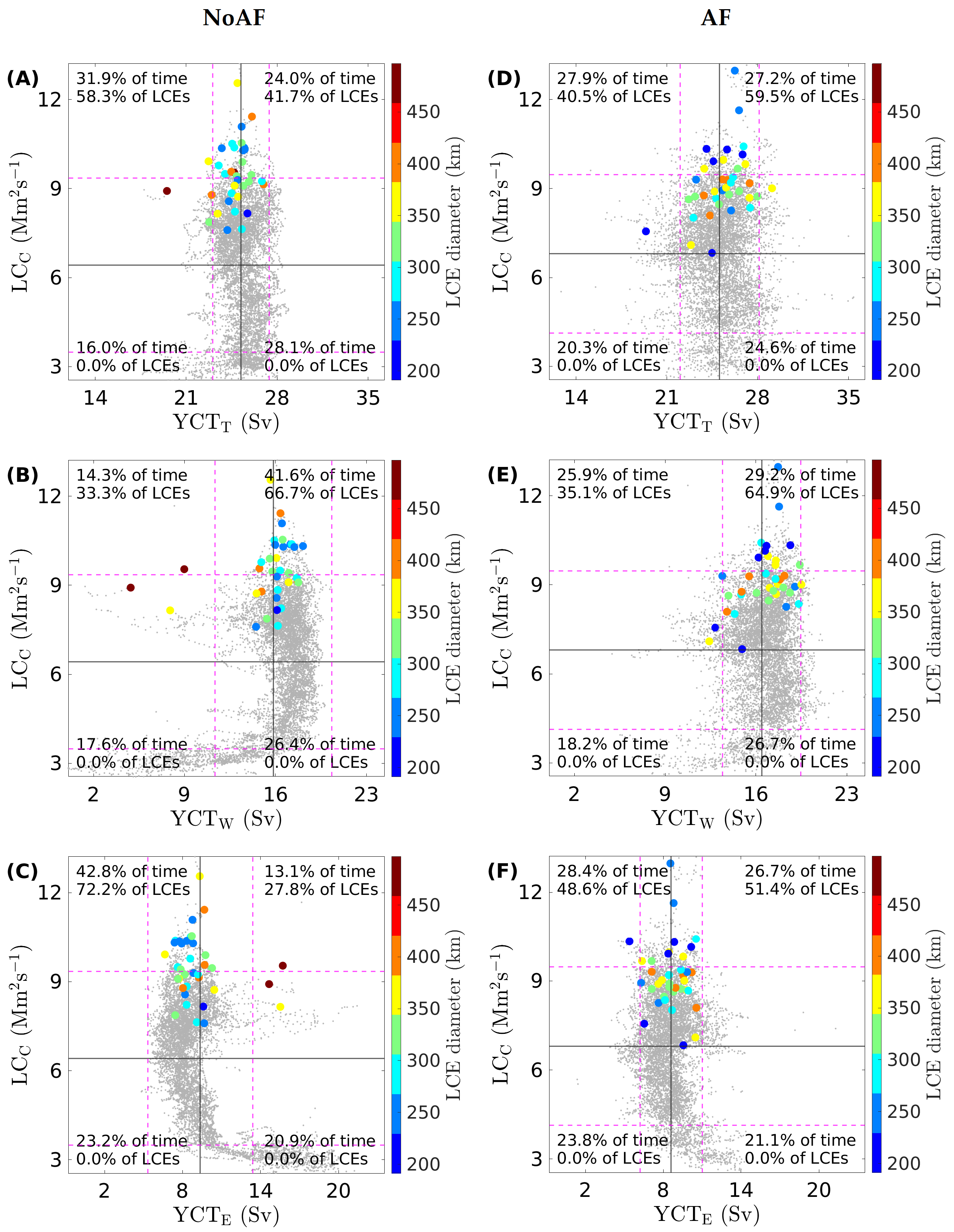}
		\caption{Scatterplots of $\mathrm{YCT_{T}}$, $\mathrm{YCT_{W}}$, and $\mathrm{YCT_{E}}$ versus $\mathrm{LC_{C}}$ for experiments NoAF and AF. The LCE separation events are indicated in colored dots according to the eddy diameter. Black solid lines indicate the mean value of each variable; pink dotted lines indicate the mean value of each variable plus and minus 1.5 their standard deviations. The lines of the mean values define quadrants in the phase space; the percentage of the data and LCE separations in each quadrant are also shown.}
		\label{Fig2}
\end{figure}

In order to analyze the short-term relationships between the YCT and LCS, scatterplots of daily time series of $\mathrm{YCT_{T}}$, $\mathrm{YCT_{W}}$, and $\mathrm{YCT_{E}}$ versus $\mathrm{LC_{C}}$ for experiments NoAF and AF are shown in Figure~\ref{Fig2}. The LCE separation events are indicated in colored dots according to the eddy diameter; black solid lines indicate the mean value of each variable; pink dotted lines indicate the mean value of each variable plus and minus 1.5 their standard deviations. The lines of the mean values define quadrants in the phase space; the percentage of the data and LCE separations in each quadrant are also shown. Three stages of the LC are distinguished: retracted ($\mathrm{LC_{C}}$ less than its mean value minus 1.5 its standard deviation), penetration ($\mathrm{LC_{C}}$ greater than its mean value minus 1.5 its standard deviation and less than its mean value), and extended ($\mathrm{LC_{C}}$ greater than its mean value).

The relationships between the LC intrusion and LCE separations in terms of YCT for experiments NoAF and AF are described below:
\begin{itemize}
\item Experiment NoAF: $\mathrm{YCT_{T}}$ (Figure~\ref{Fig2}A). In the retracted stage of the LC, $\mathrm{YCT_{T}}$ has its lowest values (as lower as 14~Sv). In the penetration stage of the LC, $\mathrm{YCT_{T}}$ is greater than its mean most of the time. In the extended stage of the LC, $\mathrm{YCT_{T}}$ is lower than its mean most of the time and the majority of LCEs separate in these conditions.

\item Experiment AF: $\mathrm{YCT_{T}}$ (Figure~\ref{Fig2}D). The structure of the retracted and penetration stages of the LC are similar to those of the experiment NoAF but with more dispersion of the data. The atmospheric forcing affects the extended stage of the LC and LCE separations. The atmospheric forcing almost balances the proportion of time that $\mathrm{YCT_{T}}$ spends at values greater and lower its mean and causes that the majority of LCEs separate when $\mathrm{YCT_{T}}$ is greater than its mean.

\item Experiment NoAF: $\mathrm{YCT_{W}}$ and $\mathrm{YCT_{E}}$ (Figures~\ref{Fig2}B and \ref{Fig2}C). An insightful description of the LC intrusion is obtained in terms of $\mathrm{YCT_{W}}$ and $\mathrm{YCT_{E}}$. The conjoint variation of $\mathrm{YCT_{W}}$ and $\mathrm{YCT_{E}}$ reflects longitudinal displacements of the YCu core. A retracted LC is associated with an eastward displacement of the YCu core ($\mathrm{YCT_{W}}$ is lower than its mean and $\mathrm{YCT_{E}}$ is greater than its mean); the lowest and greatest values of $\mathrm{YCT_{W}}$ and $\mathrm{YCT_{E}}$, respectively, are associated with events of noticeable eastward displacements of the YCu core (the $+\mathrm{YCu_{ZP}}$ events described by Higuera-Parra et al. \cite{Higuera2023}). The LC penetration is mainly associated with a westward displacement of the YCu core. In the extended stage of the LC, the YCu core has a westward displacement ($\mathrm{YCT_{W}}$ is greater than its mean and $\mathrm{YCT_{E}}$ is lower than its mean) most of the time and the majority of LCEs separate in these conditions.

\item Experiment AF: $\mathrm{YCT_{W}}$ and $\mathrm{YCT_{E}}$ (Figures~\ref{Fig2}E and \ref{Fig2}F). The characteristics of the retracted and penetration stages of the LC are similar to those of the experiment NoAF but with more dispersion of the data. As previously found, the atmospheric forcing affects the extended stage of the LC and LCE separations. The atmospheric forcing almost balances the proportion of time that $\mathrm{YCT_{W}}$ and $\mathrm{YCT_{E}}$ spend at values above and below their mean. The majority of LCEs separate when $\mathrm{YCT_{W}}$ is greater than its mean (same as in the experiment NoAF). With atmospheric forcing $\mathrm{YCT_{E}}$ is quite symmetrical with respect to its mean (a different behavior than that occurred in the experiment NoAF), suggesting a great relevance of phenomena associated to the atmospheric forcing, such as the sub and mesoscale field at the Yucatan Channel, surrounding the LC, and in the northern Gulf.
\end{itemize}

In general terms, for intrinsic ocean dynamics and with atmospheric forcing, the LC penetrates the Gulf when it is westward displaced; the atmospheric forcing strongly affects only the extended stage of the LC and LCE separations. For only ocean dynamics, the westward displacement of the LC is maintained, and the majority of the LCEs separate in this condition ($\mathrm{YCT_{W}}$ greater than its mean and $\mathrm{YCT_{E}}$ lower than its mean). When atmospheric forcing is included, the westward displacement of the LC is de-emphasized, balancing the upper meridional transport to the west and the east of $86.0^{\circ}\mathrm{W}$, and the LCEs separate at any value of $\mathrm{YCT_{E}}$, without a preference as in the case of intrinsic ocean dynamics.

The LCE separations correlate more with the longitudinal displacements of the YCu core than with the magnitudes of $\mathrm{YCT_{T}}$, $\mathrm{YCT_{W}}$, or $\mathrm{YCT_{E}}$. More details concerning the characteristics of the LC intrusion and LCE separations are obtained using $\mathrm{YCT_{W}}$ and $\mathrm{YCT_{E}}$ instead of $\mathrm{YCT_{T}}$. The above suggests that $\mathrm{YCT_{W}}$ and $\mathrm{YCT_{E}}$ play an important role in the LCS dynamics but with differentiated relationships between them. The differences in the LC intrusion and LCE separations between the experiments of intrinsic ocean dynamics and atmospheric forcing are directly related to $\mathrm{YCT_{E}}$. Thus, a meaningful description of the LCS, which captures the longitudinal displacements of the YCu core and represents the effects of the atmospheric forcing on the LCS behavior, can be done in terms of $\mathrm{YCT_{E}}$. Such a transport will be used in the subsequent analysis.

\subsection{Long-term behavior of the YCT and LCS}

\begin{figure}[p]
	\centering
		\includegraphics[angle=0, width=0.95\textwidth]{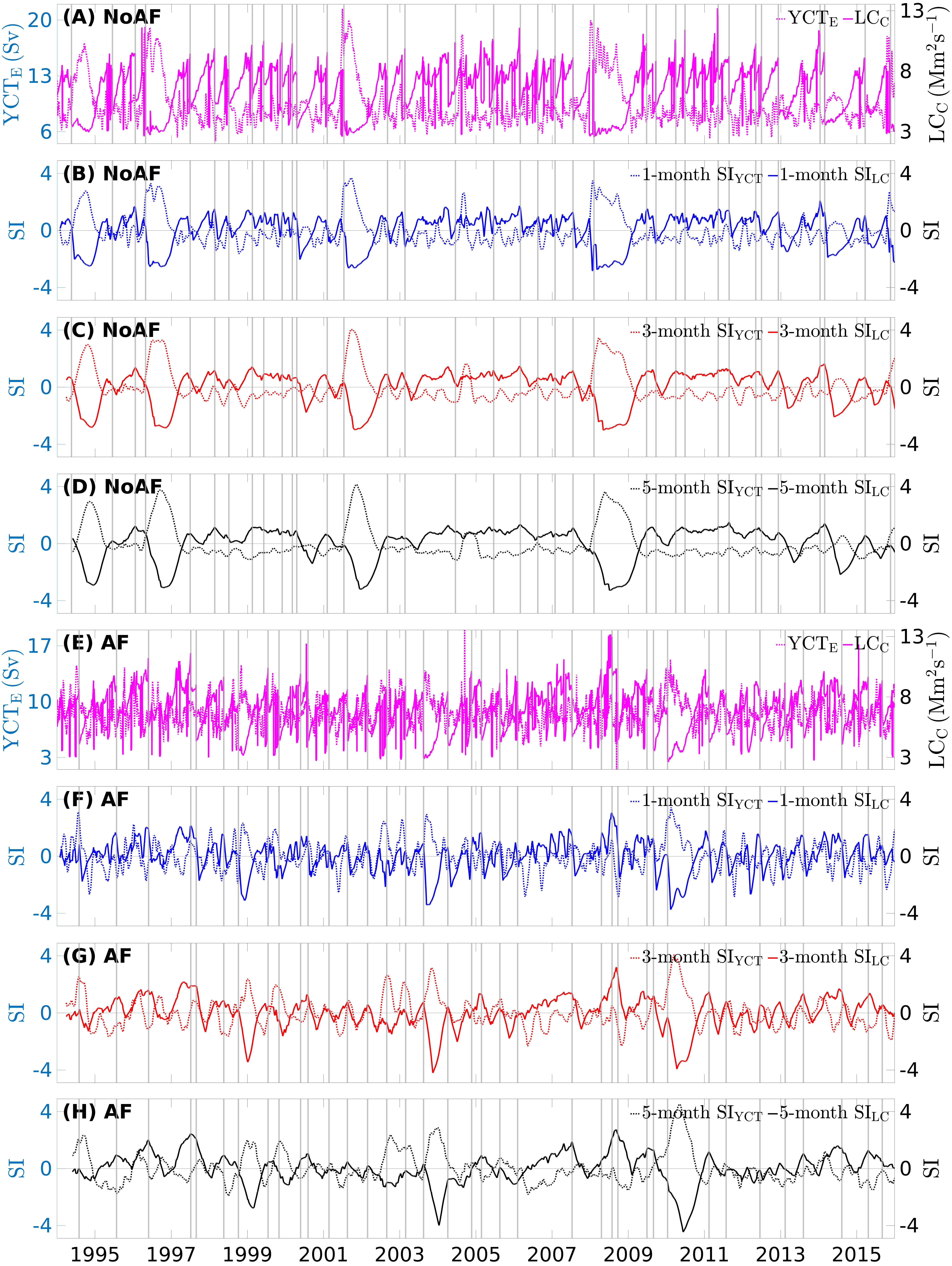}
		\caption{Time series of $\mathrm{YCT_{E}}$ and $\mathrm{LC_{C}}$ and their corresponding standardized indexes, $\mathrm{SI_{YCT}}$ and $\mathrm{SI_{LC}}$, for time scales of one, three, and five months for the experiments NoAF and AF. The dotted lines are for $\mathrm{SI_{YCT}}$ and the solid lines are for $\mathrm{SI_{LC}}$. The LCE separation events are indicated by vertical gray lines.}
		\label{Fig3}
\end{figure}

In order to analyze the long-term behavior of the YCT and LC intrusion, Figure~\ref{Fig3} shows time series of $\mathrm{YCT_{E}}$ and $\mathrm{LC_{C}}$ and their corresponding standardized indexes, $\mathrm{SI_{YCT}}$ and $\mathrm{SI_{LC}}$, for time scales of one, three, and five months for the experiments NoAF and AF; the vertical gray lines indicate LCE separations. The original daily series of $\mathrm{YCT_{E}}$ and $\mathrm{LC_{C}}$ (Figures~\ref{Fig3}A and \ref{Fig3}E) show interannual variability with a strong high-frequency signal intensified in experiment AF concerning experiment NoAF. The correlation between $\mathrm{YCT_{E}}$ and $\mathrm{LC_{C}}$ in both experiments is very complex; the most appreciable connection is that, at irregular intervals, $\mathrm{YCT_{E}}$ increases during intense and long-lasting LC retreats.

The standardized indexes help study the behavior of their associated variables at different time scales: their periods of increments and decrements (how long they last), periodicity and persistence. The standardized indexes remove the high-frequency signal of their associated variables, focusing on their low-frequency signal. The higher the index order, the more intense the variable's smoothing and the less recurrent its excursions above and below zero. The behavior of $\mathrm{SI_{YCT}}$ and $\mathrm{SI_{LC}}$ and the relationships between them at different time scales are described below (Figure~\ref{Fig3}):
\begin{itemize}
\item The standardized indexes in experiment AF (Figures~\ref{Fig3}F-\ref{Fig3}H) have a more intense high-frequency signal in comparison to those in experiment NoAF (Figures~\ref{Fig3}B-\ref{Fig3}D).

\item The occurrence of the highest and lowest values in the original daily series (Figures~\ref{Fig3}A and \ref{Fig3}E) is kept over time for the different orders of the standardized indexes (Figures~\ref{Fig3}B, \ref{Fig3}C, \ref{Fig3}D, \ref{Fig3}F, \ref{Fig3}G, and \ref{Fig3}H). However, the occurrence of such values is shifted forward in time; the higher the index order, the greater the displacement in time.

\item The larger $\mathrm{SI_{YCT}}$ and the longer it is positive, the more intense $\mathrm{YCT_{E}}$ and the longer the YCu eastward displacement. The smaller $\mathrm{SI_{YCT}}$ and the longer it is negative, the less intense $\mathrm{YCT_{E}}$ and the longer the YCu westward displacement.

\item The larger $\mathrm{SI_{LC}}$ and the longer it is positive, the more intense and longer the LC intrusion. The smaller $\mathrm{SI_{LC}}$ and the longer it is negative, the more intense and longer the LC retreat.
\end{itemize}

\begin{figure}[p]
	\centering
		\includegraphics[angle=0, width=0.95\textwidth]{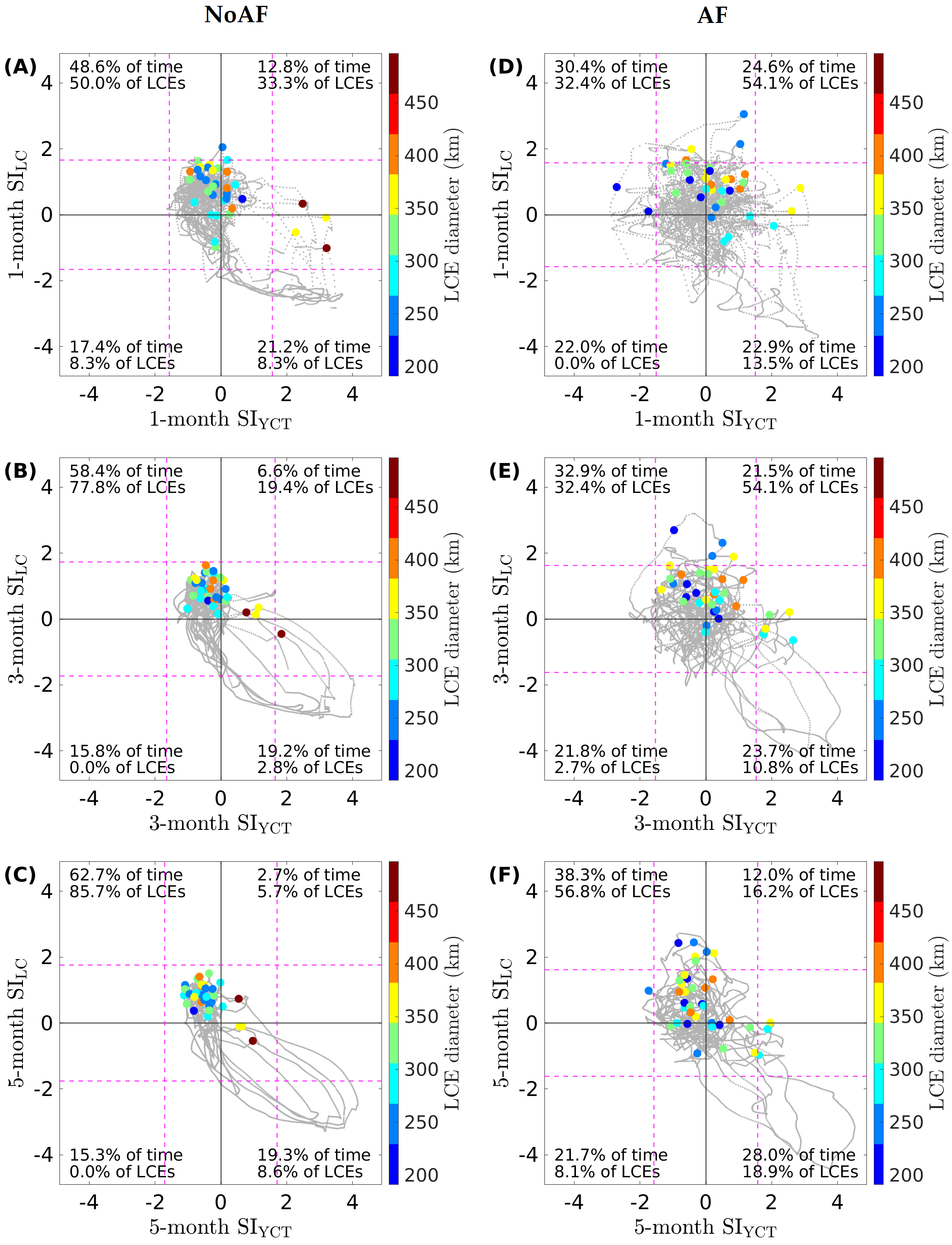}
		\caption{Scatterplots of $\mathrm{SI_{YCT}}$ versus $\mathrm{SI_{LC}}$ at different time scales for experiments NoAF and AF. The LCE separation events are indicated in colored dots according to the eddy diameter. Black solid lines indicate the zero value of each variable; pink dotted lines indicate the $\pm$1.5 standard deviation of each variable. The zero value lines define quadrants in the phase space; the percentage of the data and LCE separations in each quadrant are also shown.}
		\label{Fig4}
\end{figure}

The scatterplots between $\mathrm{SI_{YCT}}$ and $\mathrm{SI_{LC}}$ (Figure~\ref{Fig4}) provide more details regarding their relationships, including the distribution of LCE separations in terms of them. Figure~\ref{Fig4} indicates the LCE separation events in colored dots according to the eddy diameter. Black solid lines indicate the zero value of each variable; pink dotted lines indicate the $\pm$1.5 standard deviation of each variable. The three stages of the LC are also identified: retracted ($\mathrm{SI_{LC}}$ less than -1.5 its standard deviation), penetration ($\mathrm{SI_{LC}}$ greater than -1.5 its standard deviation and less than zero), and extended ($\mathrm{SI_{LC}}$ greater than zero). The zero value lines define quadrants in the phase space; the percentage of the data and LCE separations in each quadrant are also shown. Additionally, Tables~\ref{Table1} and \ref{Table2} show, for experiments NoAF and AF, the percentage of time $\mathrm{YCT_{E}}$ and $\mathrm{LC_{C}}$ spend below and above their means on different time scales, along with the corresponding percentage of separated LCEs.

\begin{table}[h] 
\begin{center}
\caption{Percentage of time $\mathrm{YCT_{E}}$ and $\mathrm{LC_{C}}$ spend below and above their means on different time scales; the percentage of separated LCEs for each case is also shown. Experiment NoAF.}\label{Table1}
\begin{tabular}{| c | c | c | c | c | c | c | c | c |}
\hline
Time & & $\mathrm{YCT_{E}}$ & & & & $\mathrm{LC_{C}}$ & & \\ \cline{2-9}
scale & $<$mean & & $>$mean & & $<$mean & & $>$ mean &  \\ \cline{2-9}
(months) & Time & LCEs & Time & LCEs & Time & LCEs & Time & LCEs \\ \hline
0 & 66.0 & 72.2 & 34.0 & 27.8 & 44.1 & 0.0 & 55.9 & 100.0 \\ \hline
1 & 66.0 & 58.3 & 34.0 & 41.6 & 38.6 & 16.7 & 61.4 & 83.3 \\ \hline
3 & 74.2 & 77.8 & 25.8 & 22.2 & 35.0 & 2.8 & 65.0 & 97.2 \\ \hline
5 & 78.0 & 85.7 & 22.0 & 14.3 & 34.6 & 8.6 & 65.4 & 91.4 \\ \hline
\end{tabular}
\end{center}
\end{table}

\begin{table}[h] 
\begin{center}
\caption{Percentage of time $\mathrm{YCT_{E}}$ and $\mathrm{LC_{C}}$ spend below and above their means on different time scales; the percentage of separated LCEs for each case is also shown. Experiment AF.}\label{Table2}
\begin{tabular}{| c | c | c | c | c | c | c | c | c |}
\hline
Time & & $\mathrm{YCT_{E}}$ & & & & $\mathrm{LC_{C}}$ & & \\ \cline{2-9}
scale & $<$mean & & $>$mean & & $<$mean & & $>$ mean &  \\ \cline{2-9}
(months) & Time & LCEs & Time & LCEs & Time & LCEs & Time & LCEs \\ \hline
0 & 52.2 & 48.6 & 47.8 & 51.4 & 44.9 & 0.0 & 55.1 & 100.0 \\ \hline
1 & 52.4 & 32.4 & 47.5 & 67.6 & 44.9 & 13.5 & 55.1 & 86.5 \\ \hline
3 & 54.7 & 35.1 & 45.2 & 64.9 & 45.6 & 13.5 & 54.4 & 86.5 \\ \hline
5 & 60.0 & 64.9 & 40.0 & 35.1 & 49.7 & 27.0 & 50.3 & 73.0 \\ \hline
\end{tabular}
\end{center}
\end{table}

We continue analyzing the long-term relationships between $\mathrm{SI_{YCT}}$ and $\mathrm{SI_{LC}}$, remembering that in the short term, the atmospheric forcing strongly affects only the extended stage of the LC and LCE separations. For intrinsic ocean dynamics (experiment NoAF), the relationships are the following:
\begin{itemize}
\item The characteristics of $\mathrm{YCT_{E}}$ being less than its mean most of the time, with most LCEs separated in this condition, are persistent from daily to longer time scales. This behavior is strongly emphasized as the time scale increases; the separation of LCEs, in terms of $\mathrm{YCT_{E}}$, is consistent for every time scale.

\item The characteristics of $\mathrm{LC_{C}}$ being larger than its mean most of the time, with all LCEs separated in this condition, are also persistent from daily to longer time scales (although with little changes as the time scale increases). This behavior is strongly emphasized as the time scale increases; the separation of LCEs, in terms of $\mathrm{LC_{C}}$, is consistent for every time scale.

\item As the time scale increases, $\mathrm{SI_{YCT}}$ and $\mathrm{SI_{LC}}$ tend to have a negative linear relationship, with the YCu core westward displaced and the LC well extended most of the time. This reflects the connection between the YCu core's position and the orientation and intrusion of the LC.
\end{itemize}

When including atmospheric forcing (experiment AF), the relationships are the following:
\begin{itemize}
\item The characteristics of $\mathrm{YCT_{E}}$ being nearly symmetrical with respect to its mean most of the time ($\mathrm{YCT_{E}}$ is lower than its mean less than 50\% of the time), with the LCEs separated at any value of $\mathrm{YCT_{E}}$, are de-emphasized from daily to longer time scales to becoming similar to those in experiment NoAF, i.e., $\mathrm{YCT_{E}}$ being less than its mean most of the time, with most LCEs separated in this condition. The separation of LCEs, in terms of $\mathrm{YCT_{E}}$, is consistent for every time scale.

\item The characteristics of $\mathrm{LC_{C}}$ being larger than its mean most of the time, with all LCEs separated in this condition, are not persistent from daily to longer time scales. As the time scale increases, $\mathrm{LC_{C}}$ tends to become symmetrical with respect to its mean, and more LCEs separate with $\mathrm{LC_{C}}$ lower than its mean, i.e., some LCEs separate with low LC intrusion or in the penetration stage of the LC ($\mathrm{SI_{LC}}<0$).

\item As the time scale increases, $\mathrm{SI_{YCT}}$ and $\mathrm{SI_{LC}}$ tend to have a negative linear relationship, with the YCu core westward displaced and the LC nearly symmetrical with respect to its mean most of the time.
\end{itemize}

The atmospheric forcing has differentiated effects on the behavior of $\mathrm{YCT_{E}}$ and the LCS. The relationship between $\mathrm{YCT_{E}}$ and the separation of LCEs is similar for intrinsic ocean dynamics and with atmospheric forcing and is consistent on every time scale. The relationship between $\mathrm{LC_{C}}$ and the separation of LCEs is different for intrinsic ocean dynamics and with atmospheric forcing, being more complex when atmospheric forcing is included. The occurrence of LCE separations with low intrusion or in the penetration stage of the LC in experiment AF suggests a significant relevance of phenomena associated with atmospheric forcing, such as the strengthened presence of sub- and mesoscale field surrounding the LC, in the Yucatan Channel, and the northern Gulf, which Higuera-Parra et al. \cite{Higuera2023} discussed. In conclusion, atmospheric forcing affects the LC intrusion and LCE separations more than $\mathrm{YCT_{E}}$.

The relationships between the behavior of the YCT and LCS are very complex. Finding statistically significant relationships between them, which help identify the LCE separations, is challenging. However, the standardized indexes of the YCT and LC metrics represent a promising alternative to describe them. They complement the information provided by the original variables and distribute the LCE separations in a suggestive manner, thereby aiding in identifying regions in the $\mathrm{SI_{YCT}}-\mathrm{SI_{LC}}$ space where LCE separations are more likely to occur.

\section{Discussion}

This work analyzed the short- and long-term behavior of the YCT, the LC intrusion, the LCE separations, their relationships, and how atmospheric forcing modifies them. It contributes to previous studies focusing on the YCT and LCS, which were based on simplified theoretical approaches \cite{Pichevin1997, Nof2005}, low- or intermediate-complexity numerical models \cite{Hurlburt1980, Hurlburt1982, Moreles2021}, and short-term observations \cite{Athie2015, Athie2020}. Using data from a long-term simulation carried out with an oceanic general circulation model made it possible to obtain realistic, robust, and statistically significant results, further enhancing our understanding of these complex phenomena. This work  confirms previously identified processes in the short term and provides results in the long term, yet to be addressed by observational analyses.

This study has provided a more comprehensive description of the LC intrusion and LCE separations by considering the eastern and western YCT, a perspective not evident when the total YCT is considered. According to Athi\'e et al. \cite{Athie2020}, the monthly time series of the western transport are significantly but weakly correlated (a Pearson's correlation coefficient of 0.42) with the LC northern intrusions when the transport leads LC extension by approximately one month. To contrast the result of Athi\'e et al. \cite{Athie2020} with the corresponding of this study, analyses of the correlation between $\mathrm{SI_{YCT}}$ and $\mathrm{SI_{LC}}$ at different orders, with $\mathrm{SI_{YCT}}$ leading $\mathrm{SI_{LC}}$ by 0, 30, and 60 days, are shown in Tables~\ref{Table3} and \ref{Table4}.

\begin{table}[h] 
\begin{center}
\caption{The Pearson's correlation between $\mathrm{SI_{YCT}}$ and $\mathrm{SI_{LC}}$ at different orders, with $\mathrm{SI_{YCT}}$ leading $\mathrm{SI_{LC}}$ by 0, 30, and 60 days. For each order, the percentage increment in the correlation with respect to the 0-day lag is shown in parentheses. Experiment NoAF.}\label{Table3}
\begin{tabular}{| c | c | c | c |}
\hline
 & 0 days & 30 days & 60 days \\ \hline
1-month & -0.660 & -0.734 & -0.733 \\
 &  & (17.3\%) & (11.2\%) \\ \hline
 3-month & -0.756 & -0.834 & -0.826 \\
 &  & (10.4\%) & (9.3\%) \\ \hline
 5-month & -0.825 & -0.877 & -0.871 \\
 &  & (6.3\%) & (5.5\%) \\ \hline
\end{tabular}
\end{center}
\end{table}

\begin{table}[h] 
\begin{center}
\caption{The Pearson's correlation between $\mathrm{SI_{YCT}}$ and $\mathrm{SI_{LC}}$ at different orders, with $\mathrm{SI_{YCT}}$ leading $\mathrm{SI_{LC}}$ by 0, 30, and 60 days. For each order, the percentage increment in the correlation with respect to the 0-day lag is shown in parentheses. Experiment AF.}\label{Table4}
\begin{tabular}{| c | c | c | c |}
\hline
 & 0 days & 30 days & 60 days \\ \hline
1-month & -0.275 & -0.325 & -0.415 \\
 &  & (18.1\%) & (51.2\%) \\ \hline
 3-month & -0.446 & -0.548 & -0.578 \\
 &  & (23.0\%) & (29.7\%) \\ \hline
 5-month & -0.620 & -0.665 & -0.653 \\
 &  & (7.3\%) & (5.4\%) \\ \hline
\end{tabular}
\end{center}
\end{table}

In agreement with Athi\'e et al. \cite{Athie2020}, YCT variations precede variations in the LC intrusion, with the highest correlation when $\mathrm{SI_{YCT}}$ leads $\mathrm{SI_{LC}}$ by 30 days in experiment NoAF. In experiment AF, the correlation is the highest for a 60-day lag; the reasons for those differences between the experiments deserve further analysis. As expected, the correlation is higher for experiment NoAF than for AF due to the complexity of the simulations. The correlation increases as the standardized index's order and the time lag increase; nonetheless, as the standardized index's order increases, the effect of an increasing time lag on the correlation is less relevant.

The LC behavior is intrinsically linked to the YCT. The description of the LCS behavior in terms of $\mathrm{YCT_{W}}$ and $\mathrm{YCT_{E}}$ showed that, despite the significant correlation between $\mathrm{SI_{YCT}}$ and $\mathrm{SI_{LC}}$ at different orders and time lags, the LC intrusion and LCE separations correlate more with the longitudinal displacements of the YCu core than with the magnitudes of $\mathrm{YCT_{T}}$, $\mathrm{YCT_{W}}$, or $\mathrm{YCT_{E}}$. These findings are consistent with those of \cite{Abascal2003, Candela2003, Sheinbaum2016, Athie2012, Athie2015, Athie2020, Higuera2023}, who documented an eastward displacement of the YCu core preceding several LCE separations. The above reinforces the validity of this research and highlights the importance of the longitudinal displacements of the YCu core on the LCS dynamics. A meaningful description of the LCS behavior can be done in terms of $\mathrm{YCT_{E}}$ since it captures the longitudinal displacements of the YCu core and accounts for the differences in the LC intrusion and LCE separations when considering intrinsic ocean dynamics and including atmospheric forcing.

By decoupling the atmospheric forcing from the Gulf hydrodynamics, it was possible to study the behavior of the YCT and LCS, considering intrinsic ocean dynamics and the relative effect of atmospheric forcing on them. This study focused on identifying the short- and long-term relationship between them, leaving for future research the analysis of the mechanisms originating variations in their behavior. Atmospheric forcing significantly influences the behavior of the YCT and LCS: it mainly affects the extended stage of the LC by creating a higher dispersion in the YCT and LC intrusion values compared to when considering intrinsic ocean dynamics. The energy associated with the short-term and random variations of atmospheric forcing is integrated to produce a response in the long term in a consistent way with the stochastic null hypothesis of climate variability proposed by Hasselmann \cite{Hasselmann1976}.

Compared to the experiment of intrinsic ocean dynamics, the LC intrusion and LCE separations have more variations when atmospheric forcing is included, suggesting a strengthened role of different phenomena driven by atmospheric forcing that affect the LCS behavior. On daily time scales, considering intrinsic ocean dynamics, the longitudinal displacements of the YCu core are mainly constrained to the west Yucatan Channel. Atmospheric forcing balances the longitudinal displacements of the YCu core, strongly affecting the LC in its extended stage. The mechanisms associated with this behavior could be related to phenomena driven by atmospheric forcing, such as vorticity anomalies from the Caribbean Sea crossing the Yucatan Channel \cite{Abascal2003, Candela2003, Sheinbaum2016, Athie2012, Athie2015, Athie2020}, as well as the presence of mesoscale eddies in the Campeche Bank, in the northern Gulf, and surrounding the LC \cite{Higuera2023, Yang2023}.

The standardized indexes are adequate for studying the long-term behavior of the YCT and LC intrusion. It was possible to associate the standardized indexes' characteristics with the YCu core's eastward displacement and the LC intrusion and analyze their propagation and persistence at different time horizons. Intrinsic ocean dynamics produces a persistent behavior of $\mathrm{YCT_{E}}$ and the LC intrusion and a consistent pattern of separation of LCEs for every time scale. On the other hand, atmospheric forcing produces differentiated effects on the YCT and LCS; its effects are more emphasized on the LC intrusion and LCE separations than on the YCT. The $\mathrm{YCT_{E}}$ behavior is persistent from daily to longer scales, but the LC behavior does not. 

The results of this work also help identify the possible expected behavior of the LCS at different time horizons as a consequence of climate change:
\begin{itemize}
\item According to the CMIP6 SSP5-8.5 scenario of climate change, for the 2040-2069 and 2070-2099 projections, the offshore wind in Mexico is expected to increase in intensity \cite{Meza2024}, which could result in an increased occurrence of LCE separations with low LC intrusion. However, such a possibility requires further analysis to confirm it since the LCS behavior depends on the wind field's intensity and structure.

\item Observations from 2003-2015 showed an eastward trend of the YCu zonal position relative to 1993-2003 \cite{Varela2018}. If this trend is maintained over long periods, it is expected: (\textit{i}) a LC intrusion with its metrics nearly symmetrical concerning their mean and (\textit{ii}) an increased occurrence of LCE separations with low or moderate LC intrusion.
\end{itemize}

\begin{figure}[h!]
	\centering
		\includegraphics[angle=0, width=0.95\textwidth]{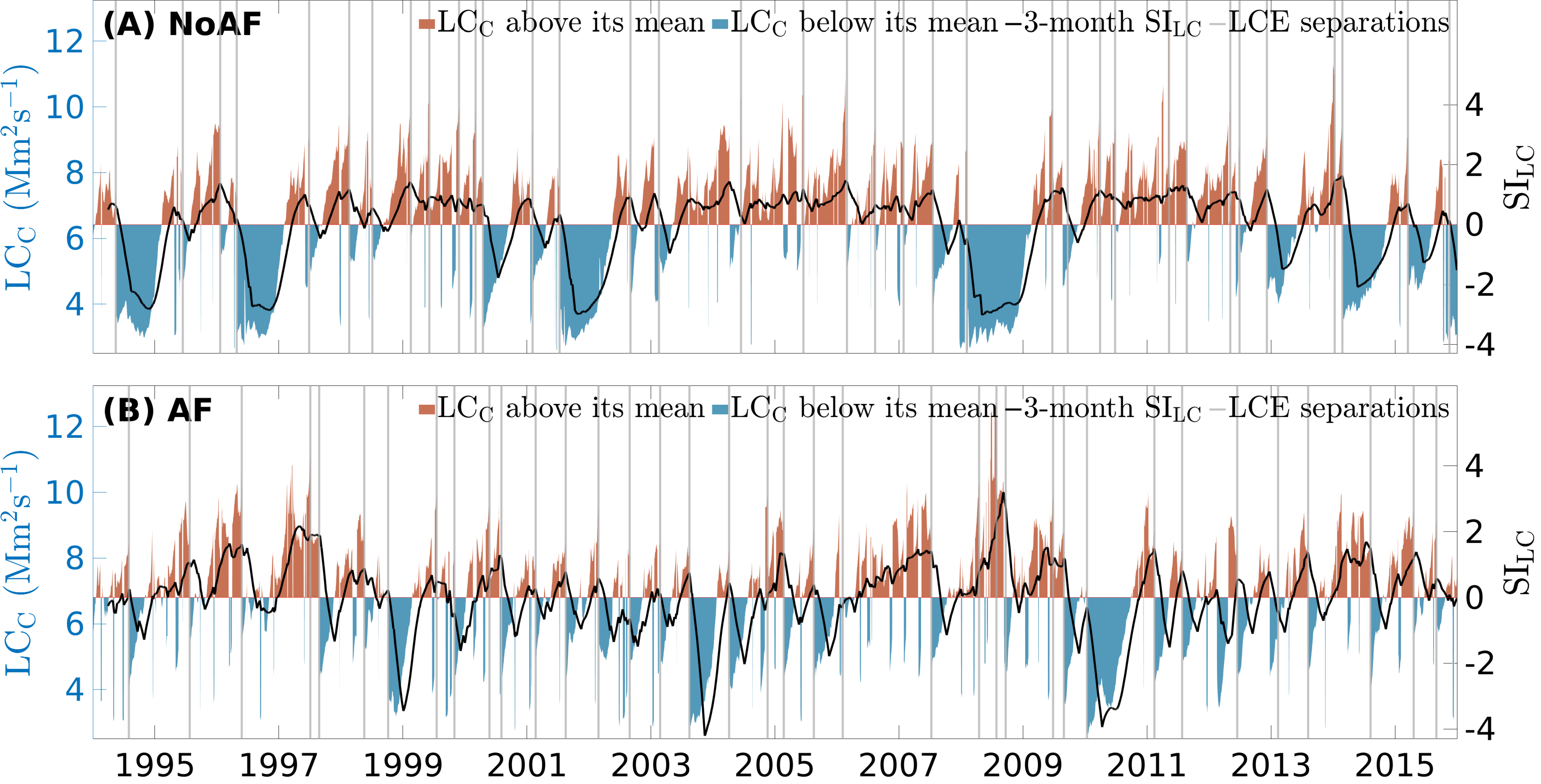}
		\caption{Time series of $\mathrm{LC_{C}}$ and the 3-month $\mathrm{SI_{LC}}$. The $\mathrm{LC_{C}}$ values above its mean are shown in red and those below its mean are shown in blue. Vertical gray lines indicate the LCE separations.}
		\label{Fig5}
\end{figure}

A future research direction of this work is associated with constructing predictive models of the LCS. Previous works have addressed this endeavor \cite{Yin2007, Zeng2015b, Chiri2019, Wang2019, Manta2023}; however, accurately forecasting the LC intrusion and LCE separation events is still a pending task. The standardized indexes of the LC metrics represent a promising alternative to improve the existing forecasts of the LC intrusion and LCE separations; $\mathrm{SI_{LC}}$ has some interesting features that can be used to achieve this aim. Figure~\ref{Fig5} shows the time series of $\mathrm{LC_{C}}$ (its values above its mean in red and below its mean in blue) and the 3-month $\mathrm{SI_{LC}}$; it also indicates the LCE separations by vertical gray lines.

$\mathrm{SI_{LC}}$ removes the fast LCE detachment events (detached LCEs that reattach to the main flow to the LC after a few days). It provides an alternative description of the LC intrusion and the LCE separations in terms of long-term variables, with direct implications for forecasting them. $\mathrm{SI_{LC}}$ represents continued periods of intrusion and retraction of the LC and its structural changes. It can identify LCE separations, but it does not provide a direct measure of the LCE diameter. The index is positive or near zero before a LCE separation and decreases afterward: the greater the decrement, the bigger the LCE, and the longer the LC retreat.

The standardized indexes, in contrast to their corresponding original LC metrics, are better for constructing a statistical forecast of the LC intrusion and LCE separations due to their more simple form and structure, which can be more easily modeled and forecast using a specific set of explicative variables (e.g., \cite{Manta2023}) or using a time series model without explicative variables (e.g., \cite{Storch1999, Brooks2019, Moreles2023}). Based on the findings of Higuera-Parra et al. \cite{Higuera2023}, who showed that the LC behavior reflects the influence of the ocean state, atmospheric forcing, and mesoscale field in the northern Gulf, it is suggested the possibility of forecasting the LC intrusion and LCE separation events using only current and past LC information.

\section{Conclusions}

This work, conducted using numerical simulations and statistical analysis, delved into two of the most significant components of the Gulf of Mexico hydrodynamics: the YCT and LCS. It analyzed the short-term relationships between them, confirming previous research results, and added evidence supporting the YCu core's longitudinal displacements as a significant mechanism for the LCE separation. In addition, this work investigated a topic that was hardly addressed: the long-term behavior of the YCT and LCS, their relationships, and the atmospheric forcing effect on them.

The study obtained statistically significant and realistic relationships between the YCT, the LC intrusion, and the LCE separation events at different time scales. The results of this research are significant for understanding the LCS behavior at different time scales and helping identify its expected behavior due to climate change, which has crucial implications for climate scientists, researchers, and policymakers. This research provides an alternative description of the LC intrusion using a standardized index that integrates the time behavior of a LC metric. This alternative could be a basis for constructing a predictive model of the LCE separations, a task proposed for future research.

There is potential for reproducing this work's analysis considering simulations incorporating interannual variability into the model's open boundary conditions or using long-term observations, which could further enhance the accuracy and reliability of our findings.

\section*{Acknowledgments}
The authors acknowledge Aurea De Jes\'us for her helpful discussions and valuable comments concerning the methodology for calculating the standardized indexes and Ana Luc\'ia de Santos for her work at the early stages of this research.

\bibliographystyle{unsrt}  
\bibliography{YCT_LCS_Moreles2024}

\end{document}